\documentclass[twocolumn,10pt,aps,prd]{revtex4-1}

\usepackage{amssymb,amsmath,amsthm}
\usepackage[pdftex]{graphicx} 
\usepackage{braket}
\usepackage[T1]{fontenc}
\usepackage[latin1]{inputenc}
\usepackage[english]{babel}

\usepackage{float}
\usepackage{braket}
\usepackage{slashed}
\usepackage{commath}
\usepackage{diagbox}
\usepackage{color}
\usepackage{bm}
\usepackage[breaklinks=true]{hyperref} 
\hypersetup{colorlinks=true,citecolor=blue,linkcolor=blue,urlcolor=blue}
\usepackage{booktabs}
\usepackage{mathrsfs}

\addto\captionsenglish{}

\addto\captionsenglish{}

\renewcommand{\epsilon}{\varepsilon}
\newcommand{\mb}[0]{\mathbf}

\renewcommand{\rm}[1]{\mathrm{#1}}

\newcommand{\p}{\partial}
\newcommand{\w}{w^{\rm{add}}}

\begin{document}
\title[]{Particle decay in expanding Friedmann-Robertson-Walker universes} 

\author{Juho Lankinen}
\email{jumila@utu.fi}
\affiliation{Turku Center for Quantum Physics, Department of Physics and Astronomy, University of Turku, Turku 20014, Finland}

\author{Iiro Vilja}
\email{vilja@utu.fi}
\affiliation{Turku Center for Quantum Physics, Department of Physics and Astronomy, University of Turku, Turku 20014, Finland}

\begin{abstract} 
The lack of energy conservation introduces new particle processes in curved spacetime that are forbidden in flat space. Therefore one has to be very cautious about using the results calculated in Minkowskian space in early universe applications. This is true for particle decay rates in particular, which need to be calculated using quantum field theory in curved spacetime. Previous studies are usually restricted to using minimal or conformal coupling for the decaying particle, while using a more general coupling would give deeper insight into particle decay. This paper presents the results we obtained for a massive particle decaying in a general power-law universe with arbitrary coupling to gravity.
We find that depending on the value of the gravitational coupling, the effect of gravitation may either strengthen or weaken the decay.
The analysis further reveals that, apart from radiation dominated universe, there are values of the coupling constant for which the decay rate is exactly Minkowskian for all universe types. Because the decay rate may be considerably modified in curved space, these issues need to be considered when doing precise cosmological calculations.
\end{abstract}

\maketitle

\section{Introduction}
For more than 100 years, Einstein's theory of general relativity has remained as the best theory of gravity relating the geometry of spacetime to its matter and energy content \citep{CMW}. Even though it has resisted quantization, a lot of insight into the fundamentals of nature has been gained through study of quantum fields propagating within the framework of classical relativity (see e.g. \citep{Birrell_Davies,Parker_Toms}). In fact quantum field theory in curved spacetime has completely changed our view of particles as in the curved spacetime the whole notion of a particle and vacuum is quite ambiguous. This led to the discovery of gravitational particle creation some 50 years ago \citep{Parker:1968,Parker:1969}. After a while, in the 1980s, Audretsch and Spangehl turned their attention to the problem of mutually interacting fields \citep{Audretsch_Spangehl:1985,Audretsch_Spangehl:1986,Audretsch_Spangehl:1987}, and provided the framework for studying relevant quantities like decay rates. However it turned out to be more challenging than in flat space.

	New phenomena are introduced into decay rate calculations when transferring into curved space. Gravitational particle creation, lack of some conservation laws, and the possibility of a field decaying into its own quanta \citep{Bros:2010,Boyanovsky:2011,Boyanovsky:2004,Boyanovsky:1997} all imply that one has to be very critical about the use of Minkowskian results in curved spacetime. 
	
	These issues have led scientists to study particle decay in more detail especially in de Sitter space \citep{Boyanovsky:2004,Boyanovsky:1997}, but also in radiation, matter and stiff matter dominated universes \citep{Tsaregorodtsev:2004,Lankinen_Vilja:2018,Lankinen_Vilja:2017b,Lotze:1989a,Lotze:1989b}. Many of these studies have been confined to some special situation, the usual assumption being that the field is either conformally or minimally coupled to gravity and the scale factor is specially chosen to correspond to a specific type of matter content. There is however no physical reason to assume that the gravitational coupling is either minimal or conformal. Indeed, there are several studies, where the coupling is different from both minimal and conformal \citep{Luo_Su:2005,Espinosa_Giudice_Riotto:2008,Herranen_etal:2015a,Hrycyna:2017,Atkins_Calmet:2013,Xianyu:2013}. However, these studies do not set significant limits on the value of gravitational coupling: either the studies are strongly model dependent or the range of the estimates are quite wide. Therefore, a more general treatment without restrictions into particular gravitational coupling or scale factor allows for a deeper insight into the interplay between these two and the decay rate greatly enhancing the understanding of the roles they play in particle decay.
	
	 In the present paper we focus on a more thorough treatment of particle decay in spatially flat Friedmann-Robertson-Walker (FRW) universes with a general power-law expansion and {\em a priori} unrestricted gravitational coupling, extending previous analyses of Refs. \citep{Lankinen_Vilja:2018,Lankinen_Vilja:2017b}. This was done using the method of added-up probabilities originally introduced in Ref. \citep{Audretsch_Spangehl:1985} without assuming any particular value for the gravitational coupling of the decaying particle. We consider a curved space generalized decay process, where a massive scalar particle interacts with two massless, conformally coupled scalars. 

	We begin with a brief introduction by giving the basic formalism of quantum field theory in curved spacetime and introduce the concept of added-up probability in Sec. \ref{sec:1}. From there, in Sec. \ref{sec:2}, we continue to calculate the transition probability and the decay rate
in spatially flat FRW-metric with a general power-law expansion. The features emerging from these calculations are presented in Sec. \ref{sec:3} and discussed in more detail in Sec. \ref{sec:4}. Finally, in Sec. \ref{sec:5} we present the conclusions. Natural units $\hbar=c=1$ are used throughout and the metric is chosen with a positive time component.

\section{Framework  and added-up probability}\label{sec:1}
\subsection{Theoretical background}
The four-dimensional spatially flat FRW spacetime is described by the metric

	\begin{align}
	ds^2=a(\eta)^2(d\eta^2-d\bm{\mathrm{x}}^2)
	\end{align}
given in conformal time $\eta$, where  $a(\eta)$ is the dimensionless scale factor. We consider a massive real scalar field $\phi$ with mass $m$ which is non-conformally coupled and a massless real scalar field $\psi$ which is conformally coupled to gravity. The Lagrangian is given by
	\begin{align}\nonumber
	\mathcal{L}=&\frac{\sqrt{-g}}{2}\big\{\p_\mu \phi \p^\mu\phi-m^2\phi^2-\xi R\phi^2+\p_\mu\psi\p^		\mu \psi-\frac{R}{6}\psi^2\big\}\\
	&+\mathcal{L}_I,
	\end{align}
where $g$ stands for the determinant of the metric, $R$ is the Ricci scalar and the coupling to gravity is controlled by the dimensionless parameter $\xi$. In four dimensions, the value $\xi=1/6$ is known as conformal coupling, while the minimal coupling is given by $\xi=0$. For the interaction term, we choose
	\begin{align}
	\mathcal{L}_I=-\sqrt{-g}\lambda \phi \psi^2,
	\end{align}
where $\lambda>0$ is the coupling constant.
The Klein-Gordon equation for the massive scalar field is
	\begin{align}\label{Klein-Gordon}
	(\square+m^2+\xi R)\phi(\eta,\mb x)=0,
	\end{align}
where $\square$ denotes the covariant d'Alembert operator.

 Because of the homogeneity of the spatial sections, the mode solutions $u_\mb k$ of Eq. \eqref{Klein-Gordon} are separable,
	\begin{align}\label{Massive_mode_solutions}
	u_\mb p(\eta,\mb x)=\frac{e^{i\mb p\cdot \mb x}}{(2\pi)^{3/2}a(\eta)}\chi_p(\eta),
	\end{align}
where $p:=|\mb p|$. For a general coupling $\xi$, the field $\chi_p$ satisfies the equation
	\begin{align}\label{KG_chi}
	\chi_p''(\eta)+\Big(p^2+a(\eta)^2m^2+a(\eta)^2(\xi-\frac{1}{6})R \Big)\chi_p(\eta)=0.
	\end{align}
Solving this equation and using the asymptotic condition, the positive mode solutions can be recognized in the standard way \cite{Birrell_Davies}. For a massless field, the corresponding mode solutions are obtained straightforwardly from the flat space solutions,
	\begin{align}\label{Massless_mode_solution}
	v_\mb{k}(\eta, \mb x)&=\frac{1}{(2\pi)^{3/2}a(\eta)}\frac{e^{i\mb{k\cdot x}-ik\eta}}{\sqrt{2k}},
	\end{align}
where $k:=|\mb k|=k^0$. 

The $S$ matrix is given as
	\begin{align}
	S=\lim_{\alpha\to 0^+}\hat{T}\exp\Big(i\int-\sqrt{-g}\lambda \phi \psi^2e^{-\alpha\eta}d^4x\Big),
	\end{align}
where $\hat{T}$ denotes the time-ordering operator. The exponential factor $e^{-\alpha\eta}$ acts as a switch off for the interaction for large times with $\alpha$ being a positive constant and called the switch-off parameter.
The perturbative expansion of the $S$ matrix for this interaction gives
	\begin{align}
	S=1-i\lambda A+\mathcal{O}(\lambda^2),
	\end{align}
where
	\begin{align}\label{A_integraali}
	A:=\lim_{\alpha\to 0^+}\int \hat{T}\phi \psi^2 e^{-\alpha\eta}\sqrt{-g}\, d^4x.
	\end{align}
Only tree level processes are considered, for which the transition amplitude is defined as 
	\begin{align}
\mathscr{A}:= \braket{\rm{out}|A|\rm{in}}.
	\end{align}

\subsection{Added-up probability}
\begin{figure}[H]
	\centering
	\includegraphics[scale=1]{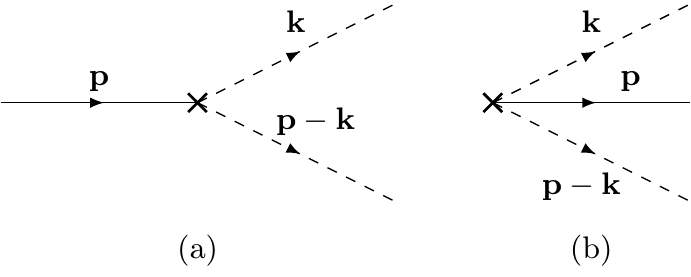}
	\caption{Production of massless particles (a) with and (b) without  proper decaying massive state. The 		solid line corresponds to the massive particle and the dashed lines to massless particles. The vertex cross  indicates gravitational influence.}\label{fig:1} 
	\end{figure}

The gravitational particle creation in curved space interferes with the process of mutual interaction, making the calculation of transition probabilities inherently difficult.
Fortunately, the added-up formalism, introduced in Ref. \cite{Audretsch_Spangehl:1985} and further investigated in Ref. \citep{Lankinen_Vilja:2018}, provides a way of calculating transition probabilities in curved space. This method relies on the fact that conformally coupled massless particles are not created from the vacuum. Therefore, a detection of a massless particle in the out-state indicates that it has solely been created or influenced by the decay process. If the out momenta is further restricted to only those massive modes fulfilling the three-momentum conservation law $\mb p= \mb k_1+\mb k_2$, one is left with what resembles closest a decay process. In the added-up formalism, the transition probability is given by
	\begin{align}\nonumber\label{eq:w_add}
	\w(\mb p,\mb k,\mb{p-k})=&\lambda^2\Big\{ \lvert \braket{\rm{out}, 1^			\psi_{\mb{k}}1^\psi_{\mb{p-k}}\lvert A \rvert 1^\phi_{\mb{p}}, \rm{out}}\rvert^2\\
	&+\lvert\braket{\rm{out},1^\phi_{\mb{-p}} 1^\psi_{\mb{k}}1^\psi_{\mb{p-k}}\lvert A \rvert 0, \rm{out}}		\rvert^2 \Big\},
	\end{align}
where $\mb k_1=\mb k$ and $\mb k_2=\mb{p-k}$. The corresponding Feynman diagrams are given in Fig. \ref{fig:1}.
\noindent The first term in \eqref{eq:w_add} corresponds to diagram (a) of Fig. \ref{fig:1}, while the second term refers to diagram (b). 
The total transition probability $w$ is obtained by summing over all the $k$-modes,
	\begin{align}\label{eq:w_tot}
	w=\sum_{\mb k} \w(\mb p,\mb k,\mb{p-k}).
	\end{align}	
A general form for the total transition probability was obtained in Ref. \citep{Lankinen_Vilja:2018},
	\begin{align}\label{eq:GeneralProb}
	w=\frac{\lambda^2}{8\pi}\int_0^T a(\eta)^2\big|\chi_{p=0}(\eta)\big|^2 d\eta,
	\end{align}	 
given in conformal time $\eta$.	Equation \eqref{eq:GeneralProb} is valid for an arbitrary scale factor $a(\eta)$ and a general field mode $\chi$ in its rest frame $\mb{p}=0$. Within this framework, we can obtain the total transition probability and the decay rate. 

\section{Transition probability and the decay rate}\label{sec:2}
We consider particle decay under a general power-law expansion $a(\eta)=b\eta^{n/2}$, with $n$ being essentially unrestricted parameter and $b$ is taken as a positive constant controlling the expansion rate of the universe. For the special choices of $n=1,2,4$, the universe is dominated by stiff matter, radiation and ordinary matter, respectively. The fact that the transition probability Eq. \eqref{eq:GeneralProb} uses only rest frame modes, enables one to solve the mode equation exactly for an arbitrary coupling $\xi$. Setting $p=0$, Eq.  \eqref{KG_chi} is given by
	\begin{align}\label{GeneralEquation}
	\chi''_{p=0}(\eta)+\big(a(\eta)^2m^2+(\xi-\frac{1}{6})Ra(\eta)^2\big)\chi_{p=0}(\eta)=0,
	\end{align}
where the Ricci scalar given by
	\begin{align}\label{RicciScalar}
	R=\frac{3n(n-2)}{2b^2\eta^{n+2}}.
	\end{align}		
The solution for Eq. \eqref{GeneralEquation} is then given in terms of Hankel functions $H^{(1)}_\alpha$ and $H^{(2)}_\alpha$ as
	\begin{align}\label{ChiSolution}\nonumber
	\chi_{p=0}(\eta)&=c_1\sqrt{\eta} H^{(1)}_\alpha\Big( \frac{2bm\eta^{(2+n)/2}}{2+n} \Big)\\
	&+c_2\sqrt{\eta} H^{(2)}_\alpha\Big( \frac{2bm\eta^{(2+n)/2}}{2+n} \Big),
	\end{align}	
where  the index
	\begin{align}\label{alpha}
	\alpha:=\frac{\sqrt{1-n(n-2)(6\xi-1)}}{2+n}
	\end{align}	
and the constants $c_1, c_2$ are determined so that the mode is correctly normalized. 

The mode solution is undefined when $n=-2$, which corresponds to de Sitter space. Even though the mode equation \eqref{GeneralEquation} can be solved for this special case, the positive energy modes cannot be identified. Because of this, the case of de Sitter space is excluded in our study.

The normalization of the mode depends on the index $\alpha$, so we have two different cases depending on whether the index is real or purely imaginary. We will treat these two cases separately and determine the restrictions they impose on $\xi$ and $n$. 

\subsection{Real index}
When the index $\alpha$ given by Eq. \eqref{alpha} is real, i.e. when $1-n(n-2)(6\xi-1)\geq 0$, the values of the coupling parameter $\xi$ are constrained to a region
	\begin{align}\label{Constraint1}
	&6\xi-1\leq \frac{1}{n(n-2)}, \ n\not\in (0,2)\\\label{Constraint2}
	&6\xi-1\geq \frac{1}{n(n-2)}, \ n\in (0,2).
	\end{align}
Using the asymptotic condition, the normalized rest frame mode is recognized as
	\begin{align}\label{eq:FieldModeRe}
	\chi_{p=0}(\eta)=\sqrt{\frac{\pi\eta}{2(2+n)}}e^{-\frac{i\pi}{4}(1-2\alpha)}H^{(2)}_\alpha \Big(\frac{2bm\eta^{(2+n)/2}}{2+n}\Big).
	\end{align}
With the change of variables $u=2bm\eta^{(2+n)/2}/(2+n)$, the total decay probability \eqref{eq:GeneralProb} takes the form
	\begin{align}\label{eq:IntegralReal}
	w_{ \mathrm{Re} }&=\frac{\lambda^2}{32 m^2}\int_0^{mt}u H^{(1)}_\alpha(u)H^{(2)}_\alpha(u)du,
	\end{align}
where $t$ denotes the standard coordinate time given by the relation $dt=a(\eta)d \eta$. 

A subtle point regarding the relationship between the standard time $t$ and the conformal time $\eta$ must be noticed; when $n<-2$, the standard coordinate time $t\in(-\infty,0)$ as $\eta\in(0,\infty)$. This situation can be remedied by making the conformal time run from $-\infty$ to $0$ in these cases, as then the coordinate time runs from $0$ to $\infty$. This has the effect of changing the scale factor to $a(\eta)^2=b^2(-\eta)^n$ and the mode solution \eqref{ChiSolution} to
	\begin{align}\nonumber
	\chi_{p=0}(\eta)&=c_1\sqrt{-\eta} H^{(1)}_\alpha\Big( \frac{-2bm(-\eta)^{(2+n)/2}}{2+n} \Big)\\
	&+c_2\sqrt{-\eta} H^{(2)}_\alpha\Big( \frac{-2bm(-\eta)^{(2+n)/2}}{2+n} \Big).
	\end{align}	
The total transition probability Eq. \eqref{eq:GeneralProb} is now given by
	\begin{align}\
	w_{\mathrm{Re}}^{(-)}=\frac{\lambda^2}{8\pi}\int_{-\infty}^0 a(\eta)^2\big|\chi_{p=0}(\eta)\big|^2 d\eta,
	\end{align}
where the cutoff is taken at the upper limit. With the change of variables to $s=-2bm(-\eta)^{(2+n)/2}/(2+n)$, the total transition probability is given by
	\begin{align}\label{eq:IntegralRealNegative}
	w_{\mathrm{Re}}^{(-)}&=\frac{\lambda^2}{32 m^2}\int_0^{mt}s H^{(1)}_\alpha(s)H^{(2)}_\alpha(s)ds,
	\end{align}
where the $t$ in the upper limit again represents the standard coordinate time. This integral is of the same form as Eq. \eqref{eq:IntegralReal}, so the analysis proceeds in the same way for all values of $n$.

The evaluation of the integral in Eq. \eqref{eq:IntegralReal} is given in the Appendix, where it is shown that the total decay probability has the exact solution
	\begin{align}\label{eq:WtotRealIndex}\nonumber
	w_{ \mathrm{Re} }&=\frac{\lambda^2}{64m^2}\{(mt)^2[J_\alpha(mt)^2-J_{\alpha-1}(mt)J_{\alpha+1}(mt)\\
	&+Y_\alpha(mt)^2-Y_{\alpha-1}(mt)Y_{\alpha+1}(mt)]-\frac{2\alpha\cot(\pi\alpha)}{\pi} \}
	\end{align}
and that the integral converges only when $|\alpha|<1$. This restricts further the allowed values of $(n,\xi)$ to 
	\begin{align}\label{Constraint3}
	\frac{(n+3)(n+1)}{n(2-n)} < 6\xi-1 \leq \frac{1}{n(n-2)},\ n\not\in (0,2)\\\label{Constraint4}
	\frac{1}{n(n-2)} \leq 6\xi-1 < \frac{(n+3)(n+1)}{n(2-n)}  ,\ n \in (0,2).
	\end{align}
When $\alpha=0$, the constant term in Eq. \eqref{eq:WtotRealIndex} cannot be used in its current form, but has to be replaced by its limiting value $2/\pi^2$.

\subsection{Imaginary index}
Purely imaginary index affects the normalization of the mode as well as the restrictions of the pair $(n,\xi)$ by reversing the inequalities \eqref{Constraint1} and \eqref{Constraint2}. Writing explicitly $\alpha=i\tilde{\alpha}$, where
	\begin{align}
	\tilde{\alpha}=\frac{\sqrt{n(n-2)(6\xi-1)-1}}{2+n}
	\end{align}
is real, the normalized positive mode is recognized to be
	\begin{align}
	\chi_{p=0}(\eta)=\sqrt{\frac{\pi\eta}{2(2+n)}}e^{-\frac{i\pi}{4}+\frac{\pi\tilde{\alpha}}{2}}H^{(2)}_{i\tilde{\alpha}} \Big(\frac{2bm\eta^{(2+n)/2}}{2+n}\Big).
	\end{align}
The only difference compared to the case of a real index is the different exponential factor with the consequence that it is not cancelled by the square of the absolute value in Eq. \eqref{eq:GeneralProb}. Performing a change of variables to $u=2bm\eta^{(2+n)/2}/(2+n)$, we are left with
	\begin{align}\label{eq:WIm1}
	w_{ \mathrm{Im} }&=\frac{\lambda^2}{32 m^2}e^{\pi\tilde{\alpha}}\int_0^{mt}u H^{(2)}_{i\tilde{\alpha}}(u)  \overline{H^{(2)}_{i\tilde{\alpha}}(u)} du.
	\end{align}
When the index on the Hankel functions is purely imaginary, there is a branch cut at negative real axis. Hence, the integrand in Eq. \eqref{eq:WIm1} is real when $u>0$. Using the following properties of the Hankel function, 
	\begin{align}
	\overline{H^{(2)}_{i\nu}(u)}=H^{(1)}_{-i\nu}(u), \ H^{(1)}_{-\nu}(u)=e^{i\nu\pi}H^{(1)}_\nu(u),
	\end{align}
the extra factor is cancelled and the total decay probability is essentially the same as for real index,
	\begin{align}\label{eq:IntegraaliIm}
	w_{ \mathrm{Im} }&=\frac{\lambda^2}{32 m^2}\int_0^{mt}u H^{(1)}_{i\tilde{\alpha}}(u)H^{(2)}_{i\tilde{\alpha}}(u)du,
	\end{align}
where a change of variables $u=2bm\eta^{(2+n)/2}/(2+n)$ has been made. The case of $n<-2$ is treated in the same manner as for the real index leading to the same form of the integral as in Eq. \eqref{eq:IntegraaliIm}. Evaluation of the integral leads to 
	\begin{align}\label{eq:WtotImaginaryIndex}\nonumber
	w_{ \mathrm{Im} }=&\frac{\lambda^2}{64m^2}\{(mt)^2[J_\alpha(mt)^2-J_{\alpha-1}(mt)J_{\alpha+1}(mt)\\\nonumber
	&+Y_\alpha(mt)^2-Y_{\alpha-1}(mt)Y_{\alpha+1}(mt)]\\
	&-\frac{2\alpha\coth(\pi\alpha)}{\pi} \},
	\end{align}
where $\alpha$ is given by \eqref{alpha}. The imaginary solution has no further restrictions to the allowed values for $\xi$ and $n$.

\subsection{Asymptotic and differential decay rates}

Besides the exact transition probabilities \eqref{eq:WtotRealIndex} and \eqref{eq:WtotImaginaryIndex}, of interest are also their asymptotic forms. Expanding the exact forms in asymptotic series, we find that the leading terms are given by
	\begin{align}
	w\sim \frac{\lambda^2}{16\pi m}\Big(t-\frac{|\alpha|\cot(\pi|\alpha|)}{2m} \Big),
	\end{align}
where we have combined the results to incorporate both real and imaginary solutions. The decay rate is obtained as usual by dividing the total probability by the time $t$. However, because of the additive term, this is more complicated than in Minkowski space. Therefore, we follow the procedure introduced in Ref. \citep{Audretsch_Spangehl:1985}, where the additive term is divided by a gravitational time $t_{grav}$. This can be defined as $t_{grav}:=t_f-t_i$, where $t_i$ indicates the time when the gravitational field begins its influence and $t_f$ its end. The mean decay rate is then
	\begin{align}\label{eq:MeanDecayRate}
	\Gamma\sim \frac{\lambda^2}{16\pi m}\Big(1-\frac{|\alpha|\cot(\pi|\alpha|)}{2mt_{grav}} \Big).
	\end{align} 	
When the conformal coupling is chosen, this more general solution reduces to those obtained in Refs. \citep{Lankinen_Vilja:2018,Lankinen_Vilja:2017b} for stiff matter, radiation and matter dominated universes. We note, that the mean decay rate \eqref{eq:MeanDecayRate} is of the form Minkowskian part plus an additive correction.

The integrand in Eqs. \eqref{eq:IntegralReal}, \eqref{eq:IntegralRealNegative} and \eqref{eq:IntegraaliIm} can be interpreted as the differential decay rate,
	\begin{align}\label{eq:DifferentialDecayRate}
	\Gamma_{\rm{diff}}=\frac{\lambda^2}{32 m} t H^{(1)}_\alpha(mt)H^{(2)}_\alpha(mt),
	\end{align}
when $t>0$ and which has no restrictions on the index values. However, if $\alpha\not\in (-1,1)$, no finite total transition probability can be defined.

\subsection{Equation of state}
We conclude this section by examining other conditions for $n$ originating from a cosmological perspective. Often the content of the Universe is described by a perfect fluid characterized by a dimensionless parameter $\omega$ equal to the ratio of its pressure $p$ to its energy $\rho$ as $\omega=p/\rho$. Moreover, in many instances the parameter $\omega$ is further restricted to $|\omega|\leq 1$, corresponding to non-phantom matter, while a hypothetical phantom matter would have an equation of state parameter corresponding to $\omega<-1$.

\begin{figure}[H]
	\centering
	\includegraphics[width=1.0\columnwidth]{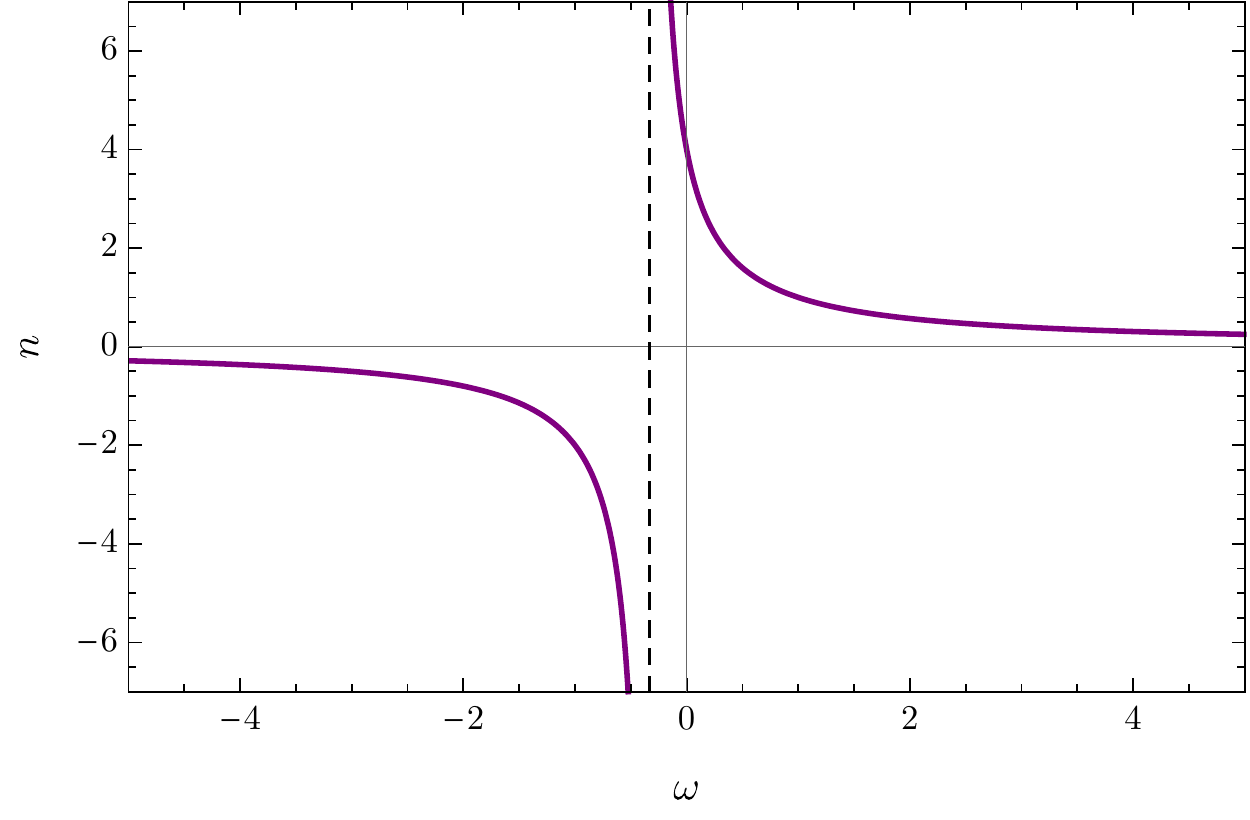}
	\caption{Values of $n$ as a function of the equation of state parameter $\omega$. The dashed line corresponds to asymptote at $\omega=-1/3$.}\label{fig:3} 
	\end{figure}

To see which values of $n$ correspond to phantom and which to non-phantom matter, we express the equation of state parameter in terms of the parameter $n$. Since the scale factor $a(t)$ in our model is proportional to $t^{n/(2+n)}$ and since the scale factor in flat FRW universe  is proportional to $t^{2/(3+3\omega)},\ \omega\neq -1$, these can be equated and solved for the equation of state parameter $\omega$. In doing so, one obtains
	\begin{align}\label{EquationofState}
	\omega=\frac{1}{3}\Big(\frac{4}{n}-1 \Big),
	\end{align}	 
valid when $n\not\in \{-2,0\}$. Because non-phantom matter is described by $|\omega|\leq 1$, one obtains from Eq. \eqref{EquationofState} that for this type of matter $n\not\in(-2,1)$. Hence, the region where $\omega<-1$ or $\omega>1$ lies in $n\in (-2,1)$. Plot of Eq. \eqref{EquationofState} is given in Fig. \ref{fig:3}.

The analysis we carried out is valid for non-phantom as well as other type of matter. The phantom energy region $n\in (-2,0)$ has received a lot of attention, see e.g. \citep{Caldwell:2002,Caldwell:2003}, but the region where $\omega>1$, corresponding to $n\in (0,1)$, is considered to be unphysical. We will mainly focus our attention on the non-phantom region.

\section{Features of the decay rate}\label{sec:3}
The generalized decay rate exhibits several interesting features and in this section we will address the most notable of them. First of them concerns the restrictions to the two parameters $(n,\xi)$ given by Eqs. \eqref{Constraint1}, \eqref{Constraint2}, \eqref{Constraint3} and \eqref{Constraint4}. These inequalities prohibit the calculation of the transition probability for certain combinations of these parameters.
We will also to discuss about the fact that the decay rate has the correct Minkowskian limit when $n=0$. This differs from our previous work \citep{Lankinen_Vilja:2017b} where we argued that the Minkowskian term in Eq. \eqref{eq:MeanDecayRate} should be taken with some caution, because the metric used did not have a well defined Minkowskian limit. This argument still holds, when $b\to 0$ is taken, but in this more general framework we find the correct limit by setting $b$ equal to unity and choosing $n=0$. 
Finally we will establish the values for the pair $(n,\xi)$, for which the relative correction term changes sign. On the asymptotic decay rate \eqref{eq:MeanDecayRate}, this has the effect of either increasing or decreasing the decay rate thereby shortening or prolonging the lifetime of the particles correspondingly. 

\subsection{The parameter space $(n,\xi)$} 
\begin{figure}[H]
	\centering
	\includegraphics[width=1.0\columnwidth]{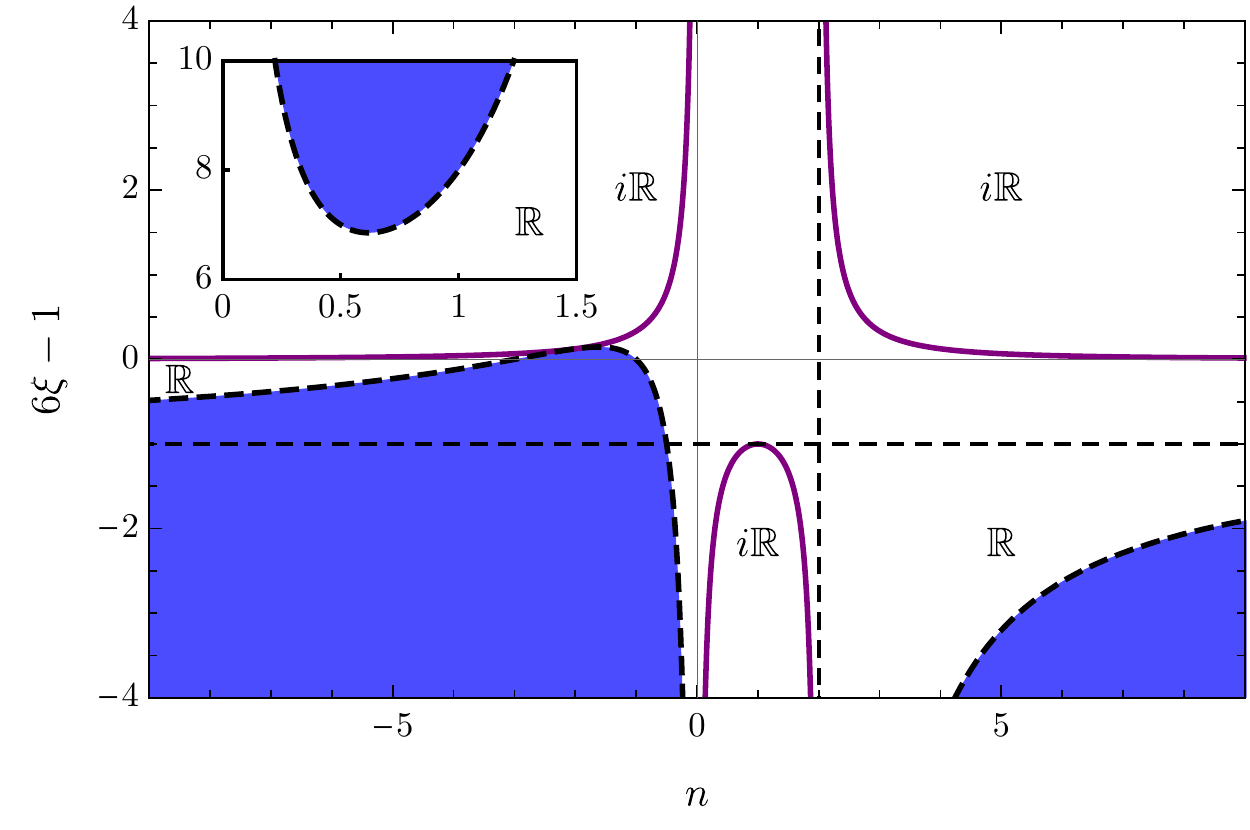}
	\caption{Allowed regions for the total transition probability. The shaded area corresponds to non-allowed values of the pair $(n,\xi)$. Regions indicated by $\mathbb{R}$ are for real index solution, while regions with $i\mathbb{R}$ are for imaginary index solutions. The dashed lines correspond to asymptotes $n=2$ and $6\xi-1=-1$.}\label{fig:2} 
	\end{figure}

The calculation of transition probability in the added-up formalism restricts the allowed pairs of the coupling $\xi$ and parameter $n$ dividing the $n\xi$-plane into regions (Fig. \ref{fig:2}).
	
For real index $\alpha$, further restrictions given by inequalities \eqref{Constraint3} and \eqref{Constraint4} eliminate regions from all quadrants of the plane. There are two regions where all values of $\xi$ are allowed, namely $n=0$ and  $n=2$. This is hardly surprising, since the Ricci scalar \eqref{RicciScalar} vanishes at these values and along with it the coupling to gravity. Moreover, there does exist one region where all values of $n$ are allowed. This band is located between the maximum and minimum points of the non-allowed region, where
	\begin{align}
	\xi\in \Big( \frac{3-\sqrt{5}}{4},\frac{3+\sqrt{5}}{4} \Big).
	\end{align}

Two values of the coupling $\xi$ are especially worthy of attention, which are the minimal coupling and the conformal coupling. These lie on the line $6\xi-1=-1$ and the horizontal axis, respectively. The conformal coupling is allowed for all values of $n$, except for a small region lying between $n\in [-3,-1]$. Even more interesting is that the boundary curves of the non-allowed regions have asymptotes at $6\xi-1=-1$ corresponding to the minimal coupling. We also observe that for non-phantom matter, the minimal coupling is allowed for all positive values of $n$ and for none when $n<-2$. Furthermore, when the universe is accelerating, i.e. $n<0$, only positive values for the coupling constant are allowed for non-phantom matter.

\subsection{Minkowskian limit and the sign of the correction term}
The chosen scale factor, $a(\eta)=b \eta^{n/2}$, has a well defined Minkowskian limit when $n=0$. The parameter $b$ can in this case be chosen equal to unity. With this choice, the parameter $\alpha$ in Eq. \eqref{alpha} is equal to $1/2$ and the field mode \eqref{eq:FieldModeRe} reduces that of flat space. Moreover, the total transition probability Eq. \eqref{eq:WtotRealIndex} reduces to
	\begin{align}
	w_{Re}=\frac{\lambda^2 t}{16\pi m}.
	\end{align}
Dividing by $t$, this corresponds to the Minkowskian decay rate. Because this is obtained from the total transition probability, it holds at all times. Thus, as would be expected, we can recover the standard Minkowskian result from the more general curved space solution making the calculation process using the added-up probability consistent. This procedure contains one caveat though, since for $n=0$ the equation of state parameter \eqref{EquationofState} is not defined. Formally we could consider this corresponding to the Minkowskian space, because for flat space the equation of state $p=\omega\rho$ has $p=\rho=0$ corresponding to zero divided by zero. The validity of this inference should be considered with great caution and in the end must be established through the use of Einstein equations.

\begin{figure}[H]
	\centering
	\includegraphics[width=1.0\columnwidth]{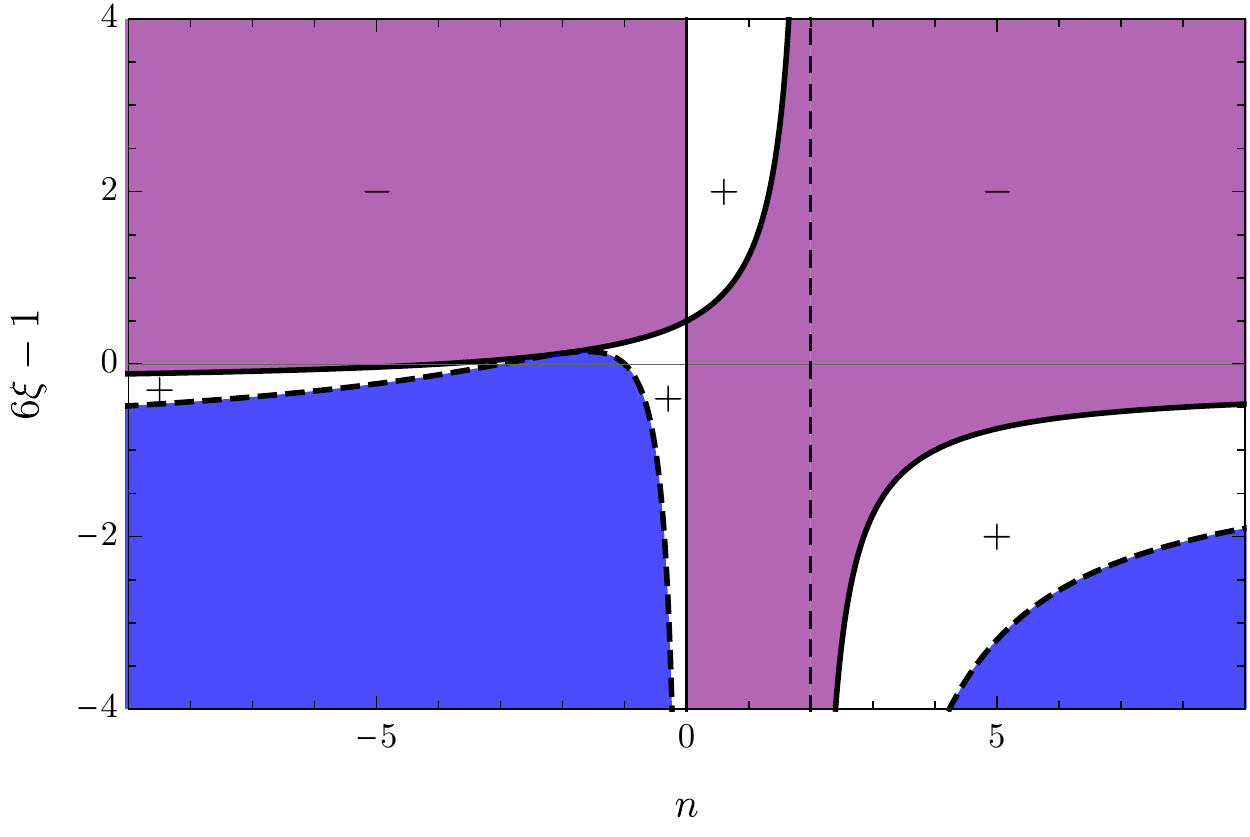}
	\caption{Plot of the Minkowskian rate curves (solid). The gravitational correction term is negative in the shaded purple region indicated by the $-$-sign. The $+$-sign indicates the areas where the contribution to the decay rate is enchancing and the shaded blue areas correspond to the values of $(n,\xi)$ where the total decay rate is not defined.}\label{fig:4} 
	\end{figure}

Since the Bessel function of the first kind is proportional to $\sin x$ for the special value of $\alpha=1/2$ and proportional to $\cos x$ for $\alpha=-1/2$, one could infer that the latter case would also reduce to the Minkowskian value. This is indeed the case and therefore the values of $\alpha=\pm 1/2$ both correspond to the Minkowskian case. This immediately implies that there are some values of $n$ and $\xi$, for which the decay rate is Minkowskian at all times, even in curved spacetime. Solving Eq. \eqref{alpha} for $\alpha=\pm 1/2$, we get two solutions. One is the already mentioned Minkowskian solution $n=0$, and the other is
	\begin{align}\label{eq:MinkowskiCurves}
	\xi(n)&=\frac{n-4}{8(n-2)}.
	\end{align}
Along this curve the decay rate is always Minkowskian (Fig. \ref{fig:4}). There is one special point on this curve; that of $n=4$, for which $\xi=0$. This means that the decay rate in a matter dominated universe for minimally coupled scalars decaying into massless conformally coupled scalars is exactly that of Minkowskian space.

\begin{figure}[H]
	\centering
	\includegraphics[width=1.0\columnwidth]{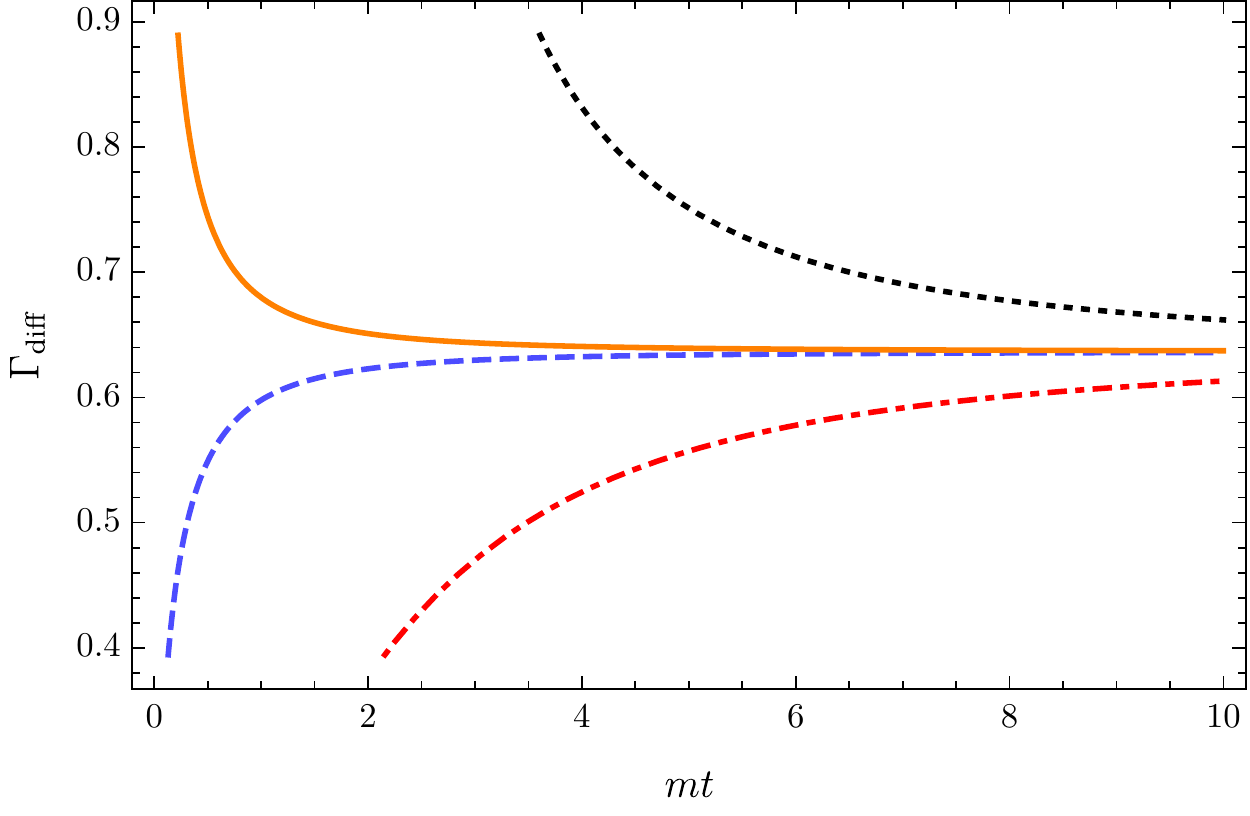}
	\caption{Plot of the differential decay rate for different values of $(n,\xi)$ in units of $\lambda^2t/(64 m)$. The dashed blue line corresponds to $(4,1/6)$, the solid orange line to $(4,-1/6)$, the dotdashed red line to $(-5,1/2)$ and the dotted black line to $(-5,-1/6)$.}\label{fig:5} 
	\end{figure}

Even though the correction term in Eq. \eqref{eq:MeanDecayRate} is negative, suggesting smaller decay rate in curved spacetime, it depends on the parameter $\alpha$ which can change signs. The change of the gravitational correction term from negative to positive happens when crossing the boundary curve \eqref{eq:MinkowskiCurves}. This curve has vertical asymptote at $n=2$ and horizontal asymptotes at $6\xi-1=\pm 1/4$. We observe, that for radiation dominated universe, the gravitational correction is always negative (Fig. \ref{fig:4}). This happens also for the conformal coupling in the non-phantom region when $n<-4$. 
Finally, we present the time evolution for the differential decay rate for different values of $(n,\xi)$ in Fig. \ref{fig:5}.  It can be seen that all curves approach the Minkowskian value asymptotically.

\section{Discussion}\label{sec:4}
	
The various features of the decaying particle raise a few questions. 
First is the fact that the transition probability can only be calculated for certain combinations of the parameters $n$ and $\xi$, thus excluding regions from the parameter space. Although perfecly clear mathematically, the appearance of these non-allowed regions presents us with a puzzling question on why they appear at all. One consideration would be if we could exclude these regions using physical arguments. For example, whether there are some constraints on the value of the non-minimal coupling $\xi$ which could be used to exclude these regions altogether.
	The value of the non-minimal coupling has attained a lot of attention recently \citep{Luo_Su:2005,Espinosa_Giudice_Riotto:2008,Herranen_etal:2015a,Hrycyna:2017,Atkins_Calmet:2013,Xianyu:2013}. Especially in Ref. \citep{Hrycyna:2017} it was argued using observational data, that negative values of the coupling would be excluded on $65\%$ confidence level. This restriction would be quite compelling to use, but the analysis in Ref. \citep{Hrycyna:2017} was done under the assumption that the model was described by a massless scalar field and is not therefore directly applicable to our situation. The value of the non-minimal coupling has also been constrained for the Higgs boson in Ref. \citep{Atkins_Calmet:2013}, where it was established that the non-minimal coupling $\xi$ should be smaller than $2.6 \times 10^{15}$. For our model, this presents no real restrictions.
	
	Because there are virtually no constraints for the non-minimal coupling which would allow us to neglect at least some of the non-allowed regions, we are led to consider whether the problem lies within the added-up formalism. At this point we must bring forth the fact that even though the mean decay rate cannot be calculated for all values of $\alpha$, the differential decay rate, identified as Eq. \eqref{eq:DifferentialDecayRate}, can. This brings us to believe, that the problem lies within the added-up probability itself, where the primary quantity calculated is the transition probability and not the decay rate, rather than there being some fundamental reason that these areas are excluded. Also, higher order corrections are not a viable option to remove these regions, since we are not doing perturbation of the coupling constant $\xi$.

The fact that the parameter $\alpha$ changes signs provides us with new insight into the relationship between particle decay rates and the gravitational coupling $\xi$. Considering first the accelerating universe, $n<0$, we can make the following observations (Fig. \ref{fig:4}). When $n=-4$, the effect of gravitation is to decrease the decay rate if the decaying field is at least conformally coupled. For values of $n$ smaller than this, the field coupling can be smaller than the conformal coupling all the way up to the asymptotic value $\xi=1/8$. For values of $\xi$ lying below the Minkowskian rate curve $\xi(n)$, the effect is to enhance the decay rate. This includes the minimal coupling, which lies in the forbidden region. Although the added-up method seems to break down in this region, it seems plausible that the effect would still be enhancing, because the differential decay rate can be calculated there.
	 Turning to the region $-4<n<0$, which includes the phantom matter as well as de Sitter regions, we see that even with the conformal coupling, the effect of gravity is to enhance the decay rate. The de Sitter point, $n=-2$, is however somewhat problematic. Even though all the curves touch at a point above the conformal coupling, where the change in the gravitational effect takes place, we cannot say anything because our solution is not defined at the point $n=-2$.
	 
		The situation is more interesting for the decelerating universe, where $n>0$. While with the conformal coupling the contribution to the decay rate is still negative, there is now greater variety in the values of $\xi$ for which the gravitational contribution is positive. For $0<n<2$, this happens when $\xi$ is sufficiently above the conformal coupling. For stiff matter dominated universe, this corresponds to $\xi=3/8$. The situation is reversed when $n>2$, because then the contribution is positive, if the gravitational coupling is sufficiently below the conformal coupling. The dividing line occurs at the matter dominated universe, where all negative values of $\xi$ give an enhancing and all positive values give decreasing contributions.

	All in all, the values of $\xi$ where the changes into decreasing or increasing decay rates happen, are quite small and seem to concentrate near the value of the conformal coupling. One exception are the values where $n$ is near the radiation dominated universe value $n=2$, where the coupling constant can be very large before any change occurs.	
	Considering then the big picture, the effect of gravitation is mostly to decrease the decay rate. This is especially plausible if the value of the non-minimal coupling is anywhere near as high as suggested in Ref. \citep{Atkins_Calmet:2013}. For the accelerating universe, this happens when the decaying particle is conformally or nearly conformally coupled to gravity. For the decelerating universe, this happens always when the particle is conformally coupled and also when it is minimally coupled, all the way up to matter dominated universe. There are also certain values for the gravitational coupling, for which the gravitation gives an increasing contribution to the decay rate. These values typically lie below the value of the conformal coupling. From all these we can infer, that the minimal coupling and the conformal coupling are somehow very different. The minimal coupling usually enhances decay rates, while the conformal couple decreases it.

The case of de Sitter space presents us with another conundrum, because it seems it cannot be treated by the added-up method when using $p=0$. The problem boils down to finding the rest frame modes for the de Sitter space. Although we can solve the mode equation \eqref{GeneralEquation} for $n=-2$, the normalized positive modes cannot be recognized because the corresponding Wronskian is zero. The rest frame mode should still be obtained from the more general solution by setting $p=0$. However, one cannot directly set $p=0$ in the commonly used Bunch-Davies mode solution \citep{Bunch_Davies:1978}, because it is not defined at this value. The decay rate might be found by not restricting to the massive particle rest frame in the added-up method, but keeping $p$ different from zero. The calculations to obtain it are, however, far from trivial. Particle decay for the same type of process in de Sitter space has nevertheless been studied using different methods of calculation \citep{Boyanovsky:1997,Boyanovsky:2004}. Unfortunately, since the added-up method cannot be used for de Sitter space, this does not allow us to compare results.

Besides these more notable features, we refer the reader to our previous works \citep{Lankinen_Vilja:2018,Lankinen_Vilja:2017b}, where we discuss in more detail about magnitude of $t_{grav}$ and a features of the asymptotic decay rates. Also the special and relevant cases of universes filled with stiff matter, radiation and ordinary matter are discussed more thoroughly in these works, when the decaying particle is conformally coupled.

\section{Conclusions and outlook}\label{sec:5}
In this paper, we have demonstrated that a more general calculation, using quantum field theory in curved spacetime, yields decay rates differing from those obtained in flat space. Depending on the strength of the gravitational coupling, the effect of gravitation is either to increase or decrease the the decay rate, thereby shortening or prolonging the lifetime of the particles correspondingly.
Even though we have addressed the most pressing need for a general decay rate analysis, there is still some issues to be resolved.

 A limitation in our calculations is the exclusion of de Sitter spacetime, which cannot be calculated using the added-up formalism with the rest frame field modes. This presents its own mathematical difficulties to be tackled on later research. The other way for calculation would of course be if the rest frame modes for a massive particle in de Sitter space could be recognized. As to the knowledge of the authors, this is yet to be done. The other major point of future studies would be the calculation of the decay rate for this exactly same process using another type of method for calculation. This could be e.g. the so called Wigner-Weisskopf method introduced in Ref. \citep{Boyanovsky:2011}. This would allow for a direct comparison between the results of two different kind of calculational methods.
 
Having these more precise decay rates allows for much more precise cosmological calculations than using just the Minkowskian approximation. These scenarios could include e.g. various baryogenesis situations or reheating after inflation. Moreover, these studies are no longer restricted to using minimal or conformal coupling for the decaying particle. Our results are not only limited to these settings but also open up the future study for a much wider class of universes filled with more exotic matter contents.

\begin{acknowledgments}
J.L would like to acknowledge the financial support from the University of Turku Graduate School (UTUGS).
\end{acknowledgments}

\appendix*

\section{Evaluation of the total transition probability integral}\label{Appendix}
The integral \eqref{eq:IntegralReal} can be written as
	\begin{align}
	I(u)=\int u H^{(1)}_\alpha(u)H^{(2)}_\alpha(u)du=\int u \big( J^2_\alpha(u)+Y^2_\alpha(u)\big )du
	\end{align}
which evaluates to
	\begin{align}\label{Appendix2}\nonumber
	I(u)=&\frac{u^2}{2}\Big(J_\alpha(u)^2-J_{\alpha-1}(u)J_{\alpha+1}(u)+Y_\alpha(u)^2\\
		&-Y_{\alpha-1}(u)Y_{\alpha+1}(u) \Big)
	\end{align}
using Eq. $5.52(2)$ from \citep{Gradstein}. The upper limit $u=mt$ is obtained by direct substitution, while the lower limit can be obtained by writing the result \eqref{Appendix2} in terms of Bessel functions of the first kind and using expansion around $u=0$. The Bessel function of the second kind can be transformed into the Bessel function of the first kind using
	\begin{align}
	Y_\nu(z)=\frac{J_\nu(z) \cos(\nu\pi)-J_{-\nu}(z)}{\sin(\nu\pi)}.
	\end{align}
The right hand side of this equation is replaced by its limiting value, when $\nu$ is integer or zero. Equation \eqref{Appendix2} can then be written as

	\begin{align}\nonumber\label{Appendix4}
I(u)=&\frac{u^2}{2\sin^2(\pi\alpha)} \Big\{ J_\alpha(u)^2 + J_{-\alpha}(u)^2 - J_{\alpha-1}(u)J_{\alpha+1}(u)\\\nonumber
&-J_{-\alpha-1}(u)J_{1-\alpha}(u)-\cos(\pi\alpha) \big[ J_{1-\alpha}(u)J_{1+\alpha}(u)\\
&+2 J_{\alpha}(u)J_{-\alpha}(u)+J_{-1-\alpha}(u)J_{\alpha-1}(u) \big] \Big\}.
	\end{align}

The product of two Bessel functions with an arbitrary index $\nu$ has a series representation, given by Eq. $9.1.14$ from \citep{Abramowitz},
	\begin{align}\nonumber
	&J_\nu(z)J_\mu(z)=\Big(\frac{z}{2}\Big)^{\nu+\mu}\\
	&\times\sum\limits_{k=0}^\infty \frac{(-1)^k\Gamma(\nu+\mu+2k+1)(z^2/4)^k}{\Gamma(\nu+\mu+1)\Gamma(\nu+\mu+1)\Gamma(\nu+\mu+k+1)k!}.
	\end{align}		
	
Expanding \eqref{Appendix4} around $u=0$, the leading terms are

	\begin{align}\nonumber
	I(u)\approx & \frac{u^2}{2\sin^2(\pi\alpha)}  \Big[ \Big(\frac{u}{2}\Big)^{2\alpha}\frac{\alpha+1}{\Gamma(2+\alpha)^2} + \Big(\frac{2}{u}\Big)^{2\alpha}\frac{1-\alpha}{\Gamma(2-\alpha)^2}\\\nonumber
	&- \Big(\frac{u}{2} \Big)^{2}\frac{\cos(\pi\alpha)}{\Gamma(2-\alpha)\Gamma(2+\alpha)}
	-\frac{\sin(2\pi\alpha)}{\pi\alpha}\\
	& + \Big(\frac{2}{u} \Big)^{2}\frac{\alpha\sin(2\pi\alpha)}{2\pi}
	 \Big].
	\end{align}			

The last term gives a constant term due to cancelling of the $u$ terms. When the index $\alpha$ is real, the lower limit of the integral converges only when  $|\alpha|<1$. These divergences are confined to the prefactors of the first two terms. When $\alpha$ is purely imaginary, there are no convergence issues and only the constant term is left. For $\alpha=0$ this has to be replaced by the limiting value $2/\pi^2$.

\end{document}